\documentclass[preprint,12pt]{elsarticle}

\usepackage{amssymb}
\usepackage[utf8]{inputenc}
\usepackage{graphicx}
\usepackage{xcolor}
\usepackage{babel}
\usepackage{alphalph}
\usepackage{natbib}
\usepackage{hyperref}
\hypersetup{
    colorlinks = true,
    anchorcolor = {blue},
}
\begin{document}

\begin{frontmatter}



\title{Machine Learning in Digital Forensics: A Systematic Literature Review}

\author[1]{Tahereh Nayerifard}
\ead{tahereh.nayerifard@mail.um.ac.ir}
%\address{Faculty of Engineering, Ferdowsi University of Mashhad, Mashhad, Iran}
\author[1]{Haleh Amintoosi}
\ead{amintoosi@um.ac.ir}
\cortext[mycorrespondingauthor]{Haleh Amintoosi}

\author[1]{Abbas Ghaemi Bafghi}
\ead{ghaemib@um.ac.ir}

\author[2]{Ali Dehghantanha}
\ead{adehghan@uoguelph.ca}

\address[1]{Faculty of Engineering, Ferdowsi University of Mashhad, Mashhad, Iran}
\address[2]{Cyber Science Lab, School of Computer Science, University of Guelph, Guelph, ON, Canada}

\begin{abstract}
Development and exploitation of technology have led to the further expansion and complexity of digital crimes. On the other hand, the growing volume of data and, subsequently, evidence is a severe challenge in digital forensics. In recent years, the application of machine learning techniques to identify and analyze evidence has been on the rise in different digital forensics domains. This paper offers a systematic literature review of the research published in major academic databases from January 2010 to December 2021 on the application of machine learning in digital forensics, which was not presented yet to the best of our knowledge as comprehensive as this. The review also identifies the domains of digital forensics and machine learning methods that have received the most attention in the previous papers and finally introduces remaining research gaps. Our findings demonstrate that image forensics has obtained the greatest benefit from using machine learning methods, compared to other forensic domains. Moreover, CNN-based models are the most important machine learning methods that are increasingly being used in digital forensics. We present a comprehensive mind map to provide a proper perspective for valuable analytical results. Furthermore, visual analysis has been conducted based on the keywords of the papers, providing different thematic relevance topics. This research will give digital forensics investigators, machine learning developers, security researchers, and enthusiasts a broad view of the application of machine learning in digital forensics.
\end{abstract}



\begin{keyword}
Digital forensic \sep Machine learning \sep Convolutional neural networks \sep Image forensics \sep Deep learning \sep SLR
\end{keyword}

\end{frontmatter}


\section{Introduction}
\label{sec:intro}
With the expanding use of digital devices and their role in human life and increasing cybercrime, digital forensics (DF) has become a significant area of research. However, there are substantial challenges in digital forensics. One of these challenges is the growing volume of data and its complexity, making the investigation process time-consuming~\cite{1, 2}. In many cases, analysis requires the classification of big data into sets that are not easy to define~\cite{3}. Another problem facing digital forensics is the diversity of data. For example, in the Internet of Things (IoT) environments, there are billions of sensors collecting various types of data, posing severe challenges in real-time cybercrime investigation cases~\cite{4}. Another essential requirement in digital forensics is accuracy and reliability in the investigation process and its results. Three factors of intelligent computing, speeding up, and reducing time, bring more reliable and accurate results in some cases~\cite{5, 6}.

In recent years, machine learning (ML) techniques in various fields such as image processing, text analysis, voice recognition, optical, and character recognition are still expanding and advancing~\cite{7}. In digital forensics, various ML techniques could gather knowledge from large volumes of digital evidence by matching conceptual models to enable data mining and knowledge discover-y~\cite{1}, and help investigators to analyze high volumes of data~\cite{8}. These methods are employed to find anomalies and identify patterns in digital forensic investigation. The automation of the investigation process in digital forensics can lead to bringing valuable aids to researchers, speeding up the process, and increasing the processing capacity~\cite{9}. Deep learning (DL) models are used in many DF domains, in adversarial image forensics~\cite{10}, image tamper detection~\cite{11}, and computer forensics~\cite{12}. These models can also be a viable solution for handling divergent data in big volumes with acceptable accuracy, e.g., analysis network traffic~\cite{13}.

 Given the importance of using ML techniques to address the digital forensics challenges and to enhance its process, in this research, a community-driven initiation has been provided to better study digital forensics and ML techniques. Identifying the ML techniques of interest in the digital forensics science community can help researchers use these techniques better and more effectively. Toward this goal, the previous investigations on digital forensics and ML have been reviewed, and new directions have been developed.
\subsection{Prior research}

To the best of our knowledge, no peer-reviewed systematic literature review has been conducted, discussing the application of ML to the problem of digital forensics explicitly. Besides, the related works have not been as comprehensive as this research thus far. The works introduced in this section have examined the relationship between DF and ML from a specific and limited view. Some papers have focused on specific techniques or applications of ML in digital forensics. Quick and Raymond Choo~\cite{1} studied some papers since 2004 about the problem of large amounts of data in digital forensics and offered solutions, including artificial intelligence (AI) and ML-based solutions. Pratama et al. \cite{5} conducted a study on digital forensics trends from the 1990s to 2014. They also had a brief review of the role of computational intelligence and its effects on digital forensics.

Faye Rona Mitchell~\cite{3} introduced AI techniques in applying pattern recognition to be used in cybersecurity and digital forensics. Some techniques such as knowledge representation, pattern recognition techniques (ML and knowledge discovery), exploratory data analysis, and knowledge refinement were briefly studied in this research. A. M. Qadir and A. Varol~\cite{8} introduced the application of using ML algorithms and techniques in digital forensics to analyze large amounts of diverse datasets to find criminal behaviours. Adam and Varol~\cite{2} studied the literature of papers between 2005 and 2019 that used classification and clustering in the process of a digital forensics investigation. They also proposed a framework for the intelligence of digital forensics. These papers only examine a specific role and application of ML in digital forensics.

Some articles have only been conducted in a specific domain of digital forensics. Kebande et al.~\cite{4} highlighted the importance of supervised ML methods in live digital forensics. They presented a framework for Emergent Configurations in IoT Environments using machine learning facilities. N. Koroniotis et al.~\cite{13} conducted a comprehensive discussion and explored the challenges of botnets and current solutions. They also studied the application of DL in network forensics and intrusion detection and its role in handling diverse data in IoT forensics as an appropriate solution. In video forensics, Abdul RehmanJaved et al. \cite{6}conducted a survey considering challenges and presenting a taxonomy of prominent video forensics products available for investigation. Al-Khateeb and Epiphaniou~\cite{14} discussed the role of ML classification techniques in an incident response methodology to improve the detection of unwanted patterns, for example, in text messages, cyberstalking, and online grooming. In 2018, Karampidis et al.~\cite{15} conducted a review of steganalysis techniques for image digital forensics. Krivchenkov et al. in~\cite{16} investigated the state of the art intelligent methods used in IoT between 2009 and 2018 and their problems in three categories of rule extraction, anomaly detection, and intrusion classification. In digital camera source identification, Jaroslaw Bernacki~\cite{17} scrutinized some methods available, including ML and DL models. Their result showed that using DL models has grown, and the CNN-based classifiers present high detection accuracy.

In a recent study conducted in 2021, Manjunatha and Patil presented a review of the DL-based passive image forensics analysis methods for detection tampering~\cite{11}. In a survey~\cite{18}, Cifuentes et al. studied using the DL methods to automate the detection of sexually explicit videos. In terms of obtaining digital evidence, Zaytsev et al.~\cite{19} revealed that AI enables a multifaceted, complicated, and objective approach to investigate crime situations and notably enhances the proof of efficiency. Jarrett and Raymond Choo~\cite{20} examined the relationship between multimedia science in three areas of Cyber Threat Intelligence, AI, and Cybercrime. They inspected the effect of AI and automation in DF on efficiency, accuracy, and cost‐reduction and explored the main automation challenges of digital forensics.
To apply the ML techniques in image manipulation detection, a comprehensive survey conducted by Norzoi et al.~\cite{10} examined the techniques available and pointed out their vulnerabilities against adversarial attacks. In another work on cybersecurity intrusion detection~\cite{21}, and file type identification~\cite{12}, the effectiveness of using ML techniques has been indicated. Shalaginov et al.~\cite{22} discussed ML techniques in static malware analysis, which can help researchers to use machine learning in malware forensics. In a systematization of knowledge (SoK), Xiaoyu Du et al.~\cite{9} studied the state-of-the-art of AI-based tools and approaches in digital forensics. In this SoK, applying AI in Data Discovery, Device Triage, Network Traffic Analysis, Handling Encrypted Data, Computer Vision, Forgery Detection, and Fingerprinting were examined, and current challenges and future directions in each field were discussed. However, this research has not been conducted as a systematic literature review.

\subsection{Research goals}
This research aims to review and identify the applications of ML in digital forensics domains. In particular, This study focuses on answering five research questions given in Table~\ref{tab:t1}.

\begin{table}[hbt]
\centering
\setlength{\arrayrulewidth}{1pt}
\scriptsize
\begin{tabular}{p{5cm}p{5cm}}
\hline
Research Questions (RQ) & Discussion\\
\hline
\textbf{RQ1}: How are publications related to ML applications in DF spread throughout the years? & To identify the trend and the progress of the subject in each year. \\
\\
\textbf{RQ2}: How is the research activity in the applications of ML in DF dispersed geographically? & To identify leading countries in this field. \\
\\
\textbf{RQ3}: What are the most popular publication venues and databases in this domain? & To identify the leading publishers and specify the extent to which conferences and journals pay attention to this subject. \\
\textbf{RQ4}: What are the most commonly used related keywords in the research, and how are they related? & Identifying the essential keywords used in the research can be helpful to categorize the trending topics and the fields receiving less attention. \\
\textbf{RQ5}: What are ML methods used in digital forensics, and in what fields? & Digital forensics use ML in various fields. Identifying the areas in which the ML techniques are utilized can help understand the ML's role in digital forensics. Because ML has a wide range of techniques, it is essential to identify the most used ones in digital forensics. \\
\hline
    \end{tabular}
    \caption{Research questions.}
    \label{tab:t1}
\end{table}

\subsection{Contributions and layout}
By considering all ML methods and DF domains, this SLR is complementary to the current research and presents the following contributions for those having a curiosity in digital forensics and ML to promote their work:

\begin{itemize}
\item We identify 608 primary studies related to ML and digital forensics up to December 2021. The results can give an excellent view to researchers in this specific field.
\item We present a meta-analysis of the state of play regarding ML methods employed to improve the digital forensics investigation process and address DF challenges.
\item We Provide visual analysis based on all 608 related articles according to the authors' selected keywords, showing thematic relevance in the last decade's research.
\item We conduct a comprehensive mind map of the relationship between ML methods and DF domains. Other researchers can use this mind map to further their work.
\item We make representations and produce guidelines to help further
work in this area.
\end{itemize}

The structure of this research is as follows: Section 2 describes the methods in which the primary studies are systematically selected for analysis. Section 3 discusses the findings related to the research questions presented earlier. Section 4 discusses the future research directions of ML application in digital forensics. Section 5 concludes the research.
\section{Research methodology}
To answer the RQs mentioned above, the SLR was conducted based on the guidance presented by Kitchenham and Charters~\cite{23}. In iterations, we desired to move via the review's planning, conducting, and reporting phases to complete the SLR evaluation.
\subsection{Selection of primary studies}
Primary studies were gathered by searching keywords in the publication's search tool or the search engine. The keywords were chosen to aid in discovering research findings that would address the research questions. We used AND and OR Boolean operators. The search terms were as follows:
\\

\textit{(”machine learning” OR ”artificial intelligence” OR "classification") AND “digital forensic”}
\\

\textit{(”neural network” OR "convolutional neural network" OR "deep neural network" OR ”deep Learning”) AND “digital forensic"}
\\

\textit{(”support vector machine” OR bayesian OR regression OR ”decision tree” OR ”k-nearest neighbor” OR supervised OR ”k-means” OR reinforcement OR "Markov" OR ”random forest”) AND “digital forensic”}
\\
\\
The platforms searched were:

\begin{itemize}
\item ACM Digital Library
\item IEEE Xplore Digital Library
\item ScienceDirect
\item SpringerLink
\end{itemize}

Depending on the search platforms, the searches were done against the title, keywords, or abstract. The final search was performed on December 10, 2021, and all studies published till that date were reviewed. The inclusion/ex-clusion criteria described in Section 2.2 were used to filter the results. The filtered results were then fed into Wohlin's~\cite{24} snowballing process. Iterations of forward and backward snowballing were done until no more papers meeting the inclusion criteria were found.

\subsection{Inclusion and exclusion criteria}
Studies can be research papers or case studies, new technical ML applicati-ons, and commentaries on developing existing digital forensics processes through the ML combination. They must be written in English and peer-reviewed. In cases where multiple versions of a study are found, the most recent version is considered. The critical inclusion and exclusion criteria are shown in Table~\ref{tab:t2}.

\begin{table}[hbt]
\centering
\setlength{\arrayrulewidth}{1pt}
\scriptsize
\begin{tabular}{p{5cm}p{5cm}}
\hline
Criteria for Inclusion & Criteria for Exclusion\\
\hline
The paper must contain information related to the ML technologies used in improving digital forensics. & Grey literature such as blogs and government documents. \\
The paper must offer empirical data related to the ML application in digital forensics. & Non-English papers \\
The paper must be peer-reviewed research published in a conference or journal. &  \\
Published between 2010 and 2021. &  \\
\hline
    \end{tabular}
    \caption{Inclusion and exclusion criteria for the primary studies.}
    \label{tab:t2}
\end{table}

\subsection{Selection results}

\begin{figure}
    \centering
    \includegraphics[scale=0.14]{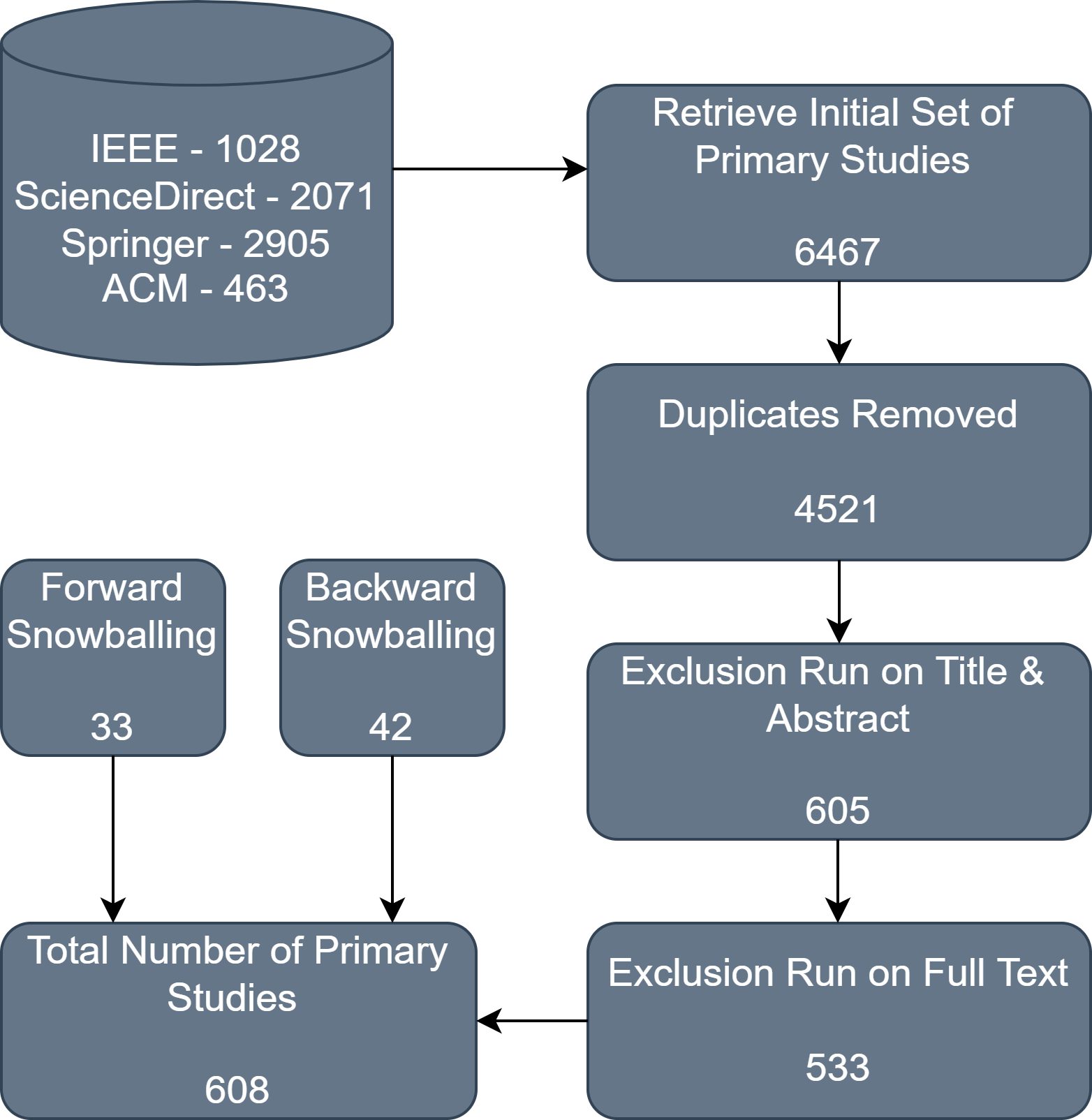}
    \caption{Attrition of papers through processing.}
    \label{fig:f1}
\end{figure}

The initial keyword searches yielded to identifying 6781 studies, which were reduced to 4521 once duplicates were removed. The remaining papers were then reviewed considering the inclusion/exclusion criteria, resulting in 605 papers being left. The 605 publications were read completely, and the inclusion/exclusion criteria were reapplied, resulting in 533 studies remaining. Snowballing was done in iterations until no more papers fulfilled the requirem-ents, resulting in 608 papers being included in the SLR. Fig.~\ref{fig:f1} displays the number of studies picked at each stage as well as the rate of attrition of articles.

\subsection{Data extraction and analysis}

The data belonging to studies that passed the quality assessment was then extracted, categorised, and saved as a spreadsheet to check for completeness and the accuracy of recording. The categories given to the data were as follows:
\begin{itemize}
\item Context data: Information about the study goal.
\item Qualitative data: Findings and conclusions provided by the authors.
\item Quantitative data: data are observed by experimentation and research when applied to the study.
\end{itemize}
To meet the goals of research questions, the data contained within the qualitative and quantitative findings categories were compiled.  In addition, a meta-analysis of those studies subjected to the final data extraction process was conducted.

\section{Discussion}
Each primary study was fully read, and relevant qualitative and quantitati-ve data was extracted provided as a mind map in answer to RQ5. All the primary studies have been focused on how ML deals with a particular problem in digital forensics. Primary keyword research shows that many papers have been published about ML techniques in digital forensics since 2006. One of the main problems in digital forensics is the large amount of evidence that makes collecting and analyzing problematic for an investigator. This high volume of data and the need to spend a very long time to find relevant evidence also increase the probability of human error. The study results show that the ML techniques can be well used in response to this problem and increase the accuracy and speed of the investigation process in data collection, inference, and analysis. The proposed ML-based methods present outstanding performance to improve the accuracy and reduce the error rate. However, limitations and challenges often relate to the ML's nature, such as the need for adequate and appropriate training data and samples~\cite{120, 310, 355, 365, 366, 409, 463}, the selection of features~\cite{233, 385, 445, 516}, the number of features~\cite{169, 172, 366}, the massive number of parameters, and the determination of optimal values~\cite{48}.

\subsection{Quantitative data}
This section provides a quantitative analysis of the set of studies resulting from the primary research. In particular, the research questions RQ1 to RQ3 were addressed by analyzing the number of publications related to the use of ML in DF over the years, the geographical distribution of these studies, and the favourite publication venues.

\subsubsection{RQ1: Spread of Publications Throughout the Years}
\begin{figure}
    \centering
    \includegraphics[width=\textwidth]{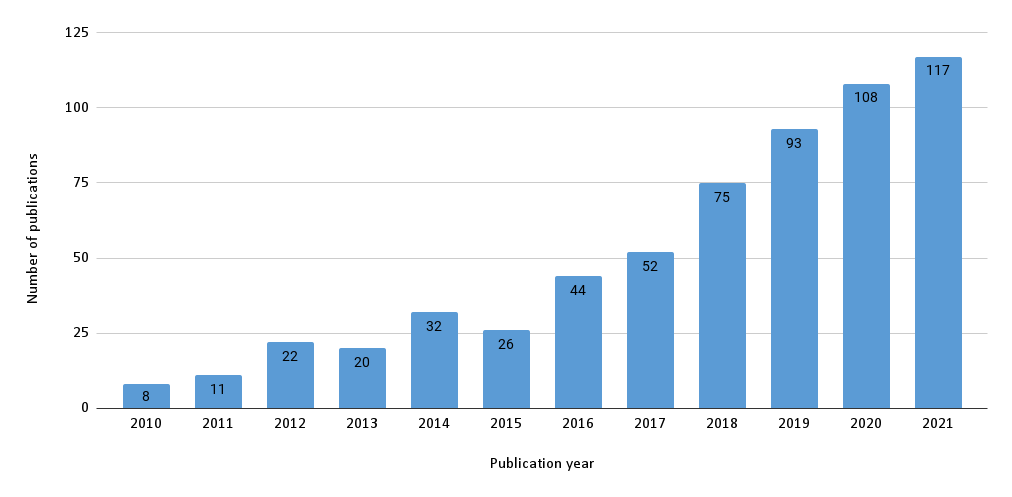}
    \caption{Publication year}
    \label{fig:f2}
\end{figure}

Fig.~\ref{fig:f2} indicates the number of publications between 2010 and 2021. Altho-ugh the research in ML application in digital forensics has grown between 2010 and 2015, it has been more impressive since 2016. A sharp increase observed from 2016 to 2021 shows that ML application in DF has been at the centre of interest of the research community.

\subsubsection{RQ2: Geographical Distribution of using ML in DF Research}
\begin{figure}
    \centering
    \includegraphics[scale=0.15]{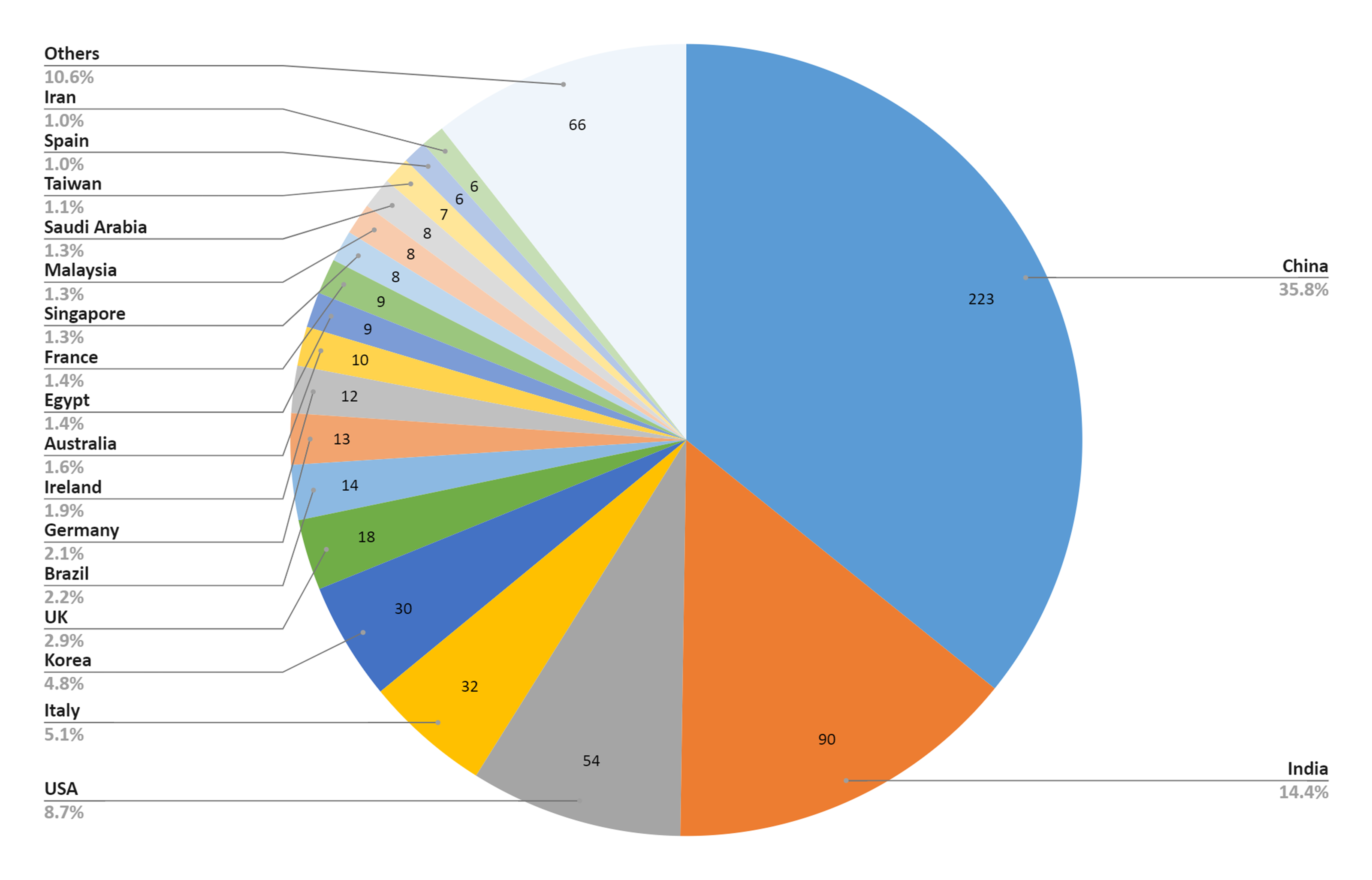}
    \caption{Demographic: geographical distribution of research activity based on the first author’s country of affiliation.}
    \label{fig:f3}
\end{figure}

\begin{figure}
    \centering
    \includegraphics[scale=0.6]{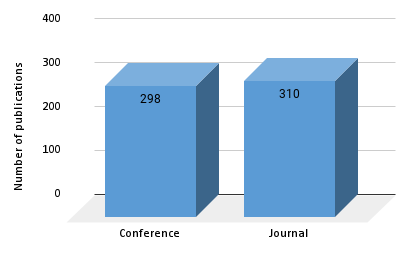}
    \caption{The popularity of different venue types.}
    \label{fig:f4}
\end{figure}

\begin{figure}
    \centering
    \includegraphics[scale=0.6]{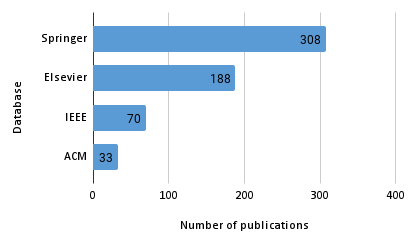}
    \caption{The number of publications in different databases.}
    \label{fig:f5}
\end{figure}

The geographical distribution of the research activity has been shown in Fig.~\ref{fig:f3}. The data were obtained by extracting the first author's affiliation with the desired studies. China, India, and the USA are the three major contributors. This can be due to the size of their industries and the importance of research in this field. However, about 18\% of the studies emerge from Europe, indicating that this topic is less intriguing. The 21 countries in the 'others' group with five or fewer publications in this field are: Greece, Turkey, Norway, Pakistan, Japan, Russia, Austria, Vietnam,  Netherlands, Canada, Bangladesh, Romania, Poland, United Arab Emirates, Colombia, Hong Kong, Jordan, Lithuania, Portugal, South Africa, and Estonia.
The papers were classified based on their publication's venue type and database, as shown in Figs.~\ref{fig:f4} and ~\ref{fig:f5}, respectively. As can be seen, conference proceedings are more active in publishing papers compared to journals. Although the Springer database has the highest number of publications, most journal papers belong to Elsevier, with about 47\% of all.

\subsection {Quality data}
In this section, a quality analysis of the primary studies was provided to answer RQ4 and RQ5.

\subsubsection{RQ4: Keyword network: analysis for identifying research areas}
\begin{figure}
    \centering
    \includegraphics[width=\textwidth]{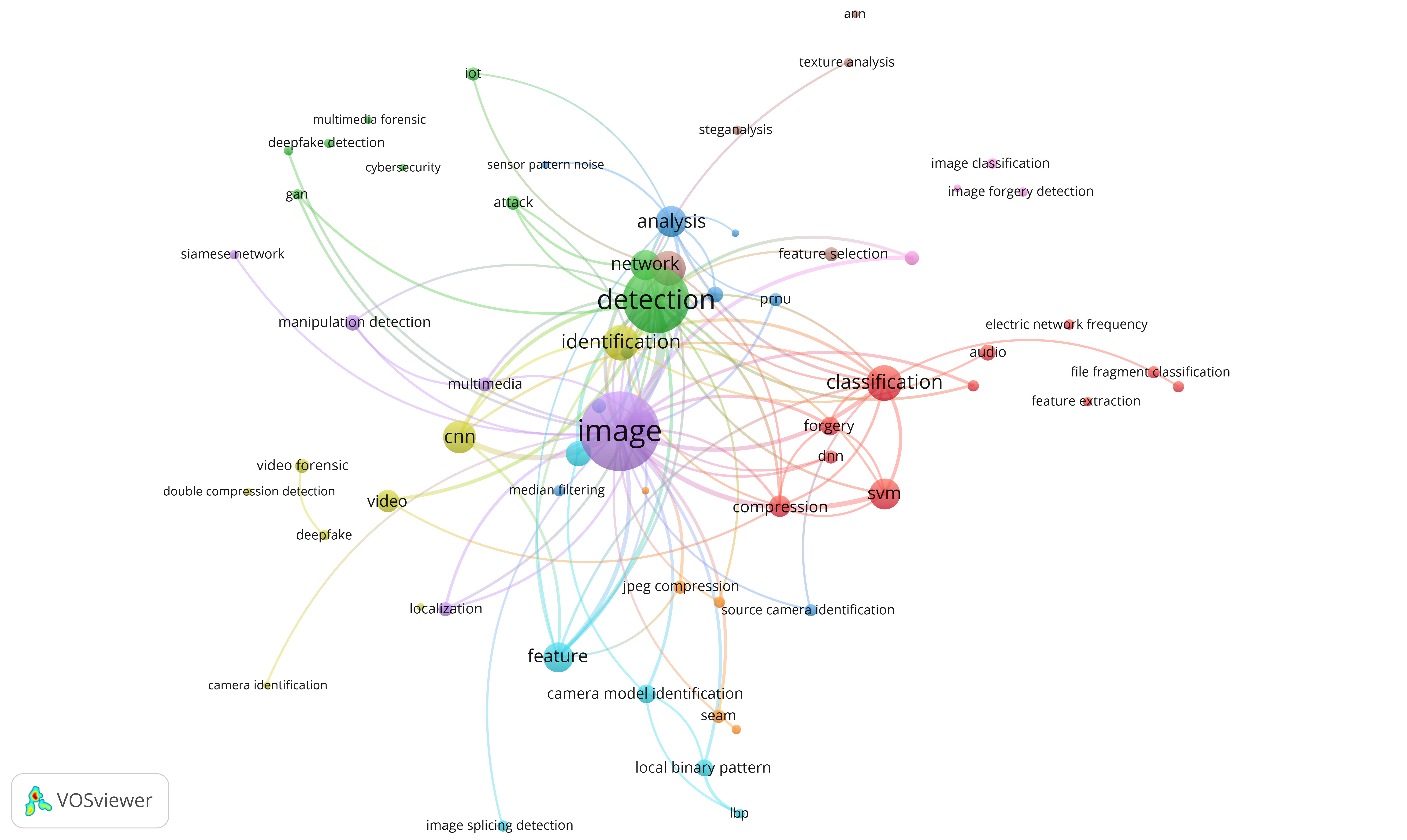}
    \caption{Map of a co-occurrence network of authors’ keywords.}
    \label{fig:f6}
\end{figure}

To summarize the common topics amongst the selected primary papers, keywords were analysed across all 608 papers based on authors' keywords. Table~\ref{tab:t3} shows the significant words that have been repeated more than 20 times along with their number of repetitions, and Fig.~\ref{fig:f6} shows their relations. For this aim, a keyword network analysis was conducted using VOSviewer~\cite{25}. At first, the social network map of the co-occurrence matrix was obtained. Fig. 6 shows a network based on the repetition of authors' keywords in the literature. Based on the similarity of keywords in topics, a classification of the keywords with the most thematic relevance is displayed in different colours. It could intuitively reveal the relationship of research themes of using machine learning in digital forensics. The size of the node's font indicates the frequency of keywords: the higher the frequency causes the larger the node's font size. The thickness of the line is relevant to the closeness of connections between two keywords. The settings in VOSviewer are based on the number of occurrences of a term=5 and Binary counting with Max. length=30, Max. lines=100 and weights= Occurrences.

\begin{table}[hbt]
\centering
\setlength{\arrayrulewidth}{1pt}
\scriptsize
\begin{tabular}{p{2.5cm}p{2.5cm}}
\hline
Keywords & Count\\
\hline
Image & 347 \\
Detection& 210 \\
CNN & 139 \\
Forensic(s) & 122\\
Identification & 91 \\
DL & 91 \\
Forgery & 81 \\
SVM & 71\\
classification & 69\\
Video & 61\\
Compression & 58\\
Jpeg & 43\\
Camera & 42 \\
Splicing & 41\\
Audio & 31 \\
Computer & 28\\
Multimedia & 27 \\
\hline
    \end{tabular}
    \caption{Counts of the significant authors' keywords in the primary studies}
    \label{tab:t3}
\end{table}

\begin{figure}
    \centering
    \includegraphics[width=\textwidth]{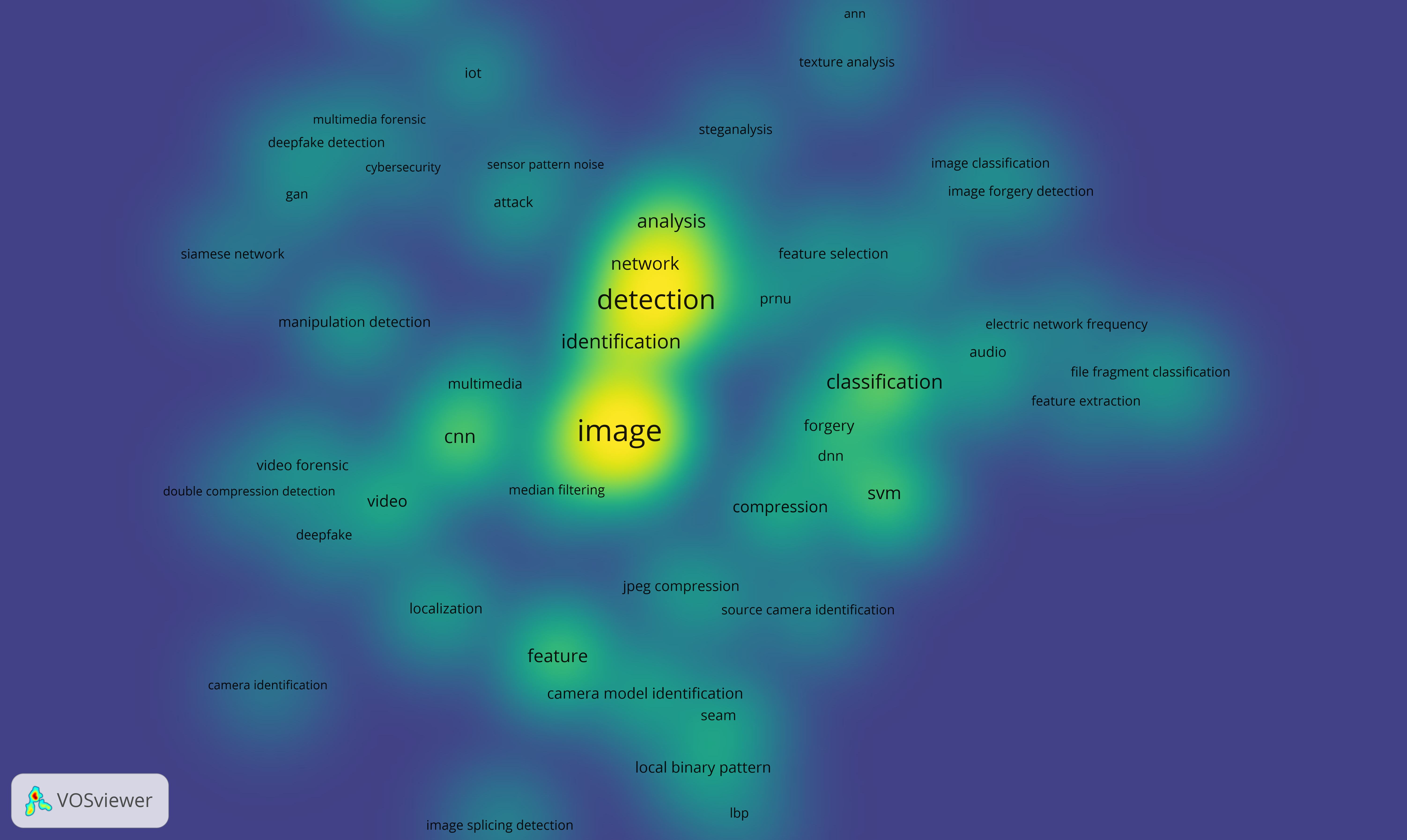}
    \caption{Density visualization of authors’ keywords.}
    \label{fig:f7}
\end{figure}

A graph of density visualization is illustrated in Fig.~\ref{fig:f7} with a kernel size of 1.5 (default) to acquire more information on these keywords. In the item density visualization, the items indicated by their labels are similar to the network visualization. The colour of each point on the map is determined by the density of its items. In the item density visualization, items are represented by their labels, similar to the network visualization. By default, the range of colours is blue to green to yellow. Larger number of items in the neighbourhood of a point and higher weights of neighbouring items result in the colour of the point closer to yellow. On the other hand, the smaller values are blue~\cite{25}.

The goal of this research question is two folds. Firstly, the aim is to determine which digital forensics domains successfully used the ML methods, leading to their advancement. Secondly, the aim is to specify the ML techniqu-es with the most capabilities to be used in digital forensics and discover the DF domains which have used each of these ML techniques. Notice that the last perspective is different from the first one.

\begin{figure}
    \centering
    \includegraphics[scale=0.2]{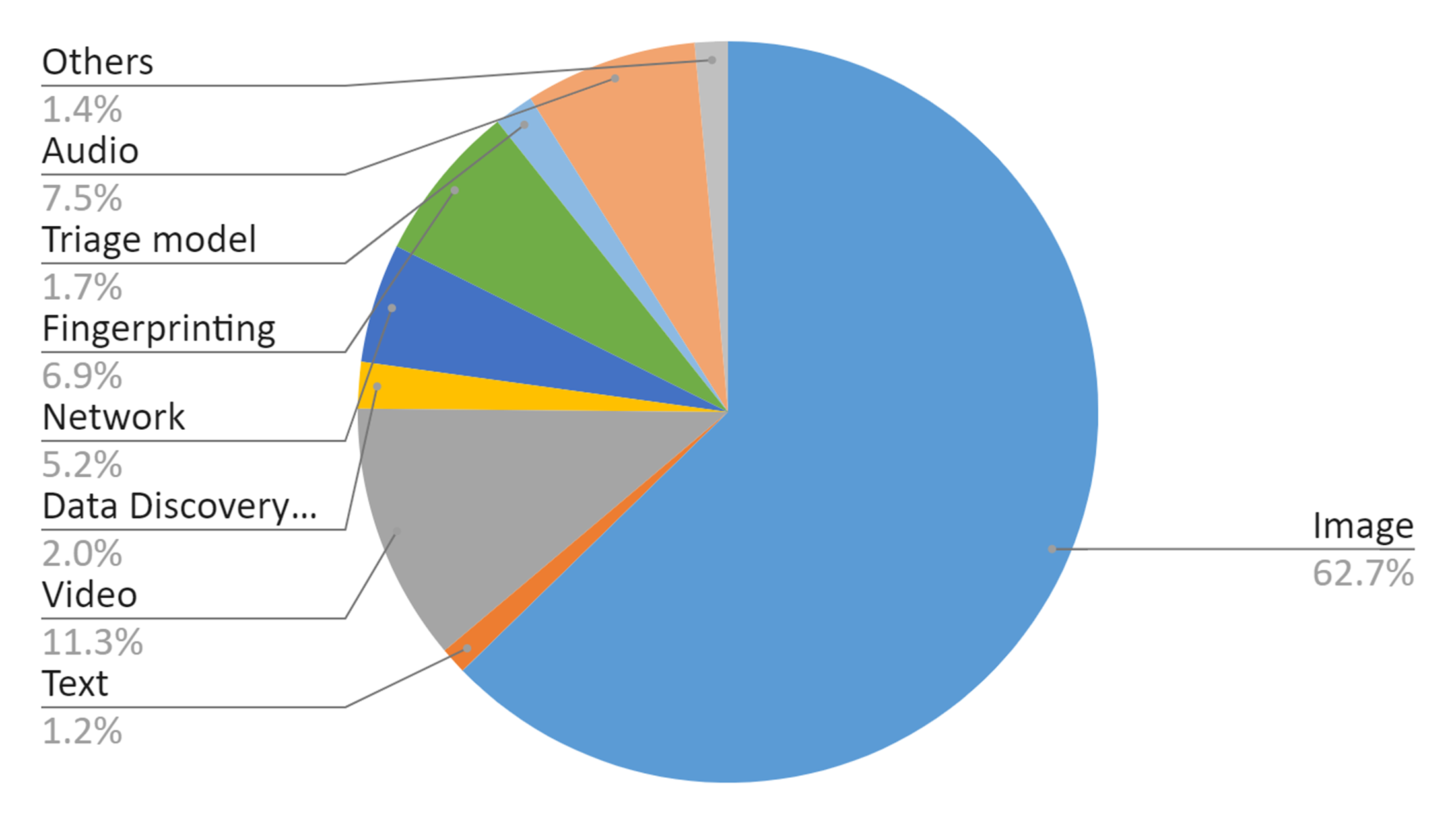}
    \caption{DF domains using ML methods (Based on authors' keywords)}
    \label{fig:f8}
\end{figure}

To answer this question, the categorization of digital forensics domains presented in~\cite{9} was considered. These categories are Data Discovery and Recovery, Fingerprinting, Multimedia Forensics (Image, Video, Audio, Text), Network, and Triage mode as shown in Fig.~\ref{fig:f8}. The statistical results of this section are based on the authors' keywords. It should be noted that the papers related to Electromagnetic side-channel analysis were included in the network category due to their relevance to the IoT and malware detection. However, to better identify the relationship between ML techniques and digital forensics domains, a comprehensive mind map is presented in Fig.~\ref{fig:f10} to enable researchers to identify the field of study of papers more accurately. This mind map is based on the paper's context and not just keywords. Due to the existence of many works in the multimedia domain, and for more clarity of the presented results, this domain is considered with its subsets in Fig.~\ref{fig:f8}. As can be seen, image forensics has been at the centre of interest among other domains. A statistical analysis of the papers revealed that about 62\% of the works were related to image forensics. Video with about 11\%, followed by audio and fingerprinting domains with about 7\%, have been exciting topics to researchers. The 'others' group of DF domains consists of three forensics phrases written by the authors, namely, sensor forensics, computational forensics, and location forensics.

\begin{figure}
    \centering
    \includegraphics[scale=0.25]{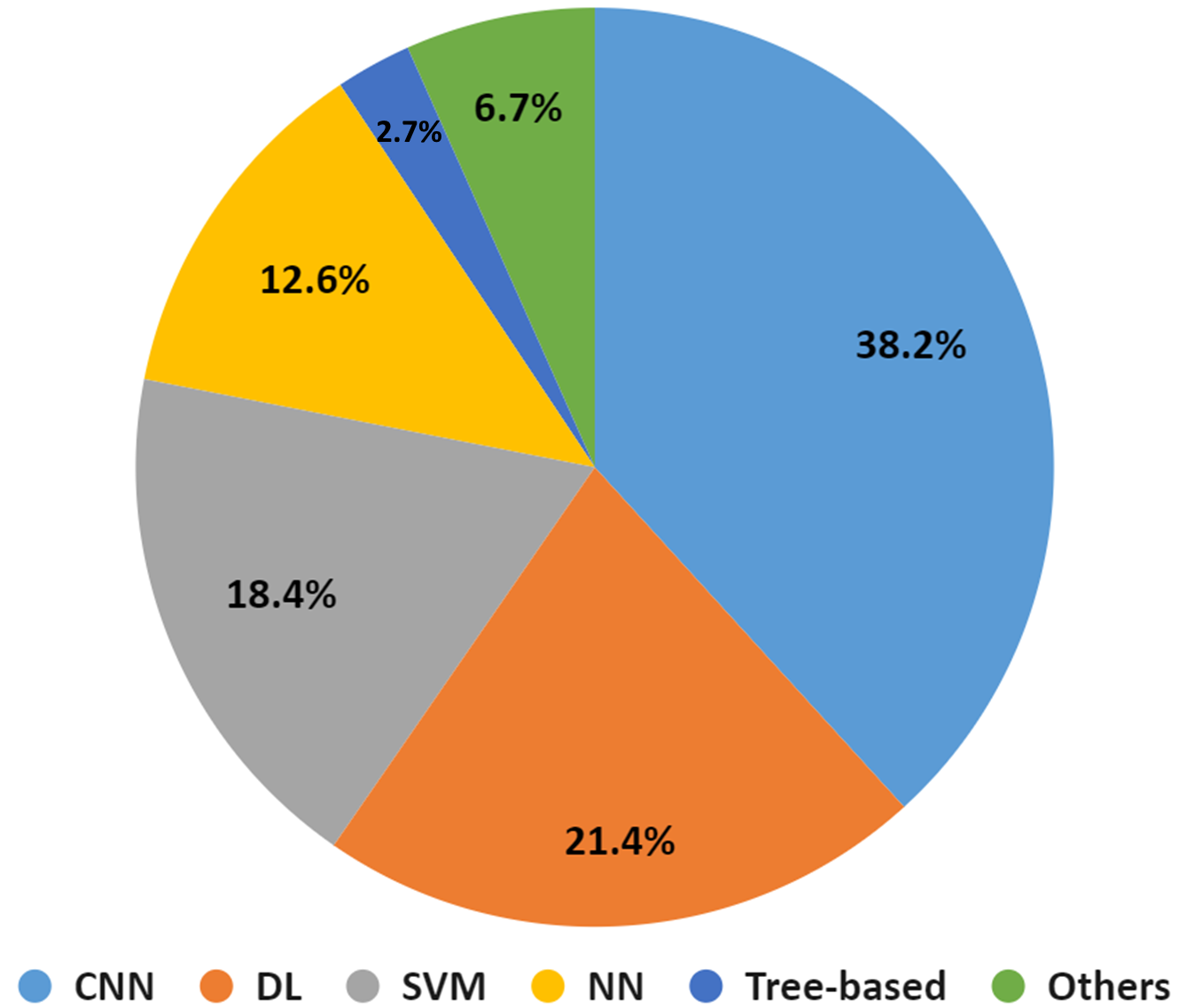}
    \caption{The main ML methods used in DF  (Based on authors' keywords)}
    \label{fig:f9}
\end{figure}

\begin{figure}
    \centering
    \includegraphics[scale=0.4]{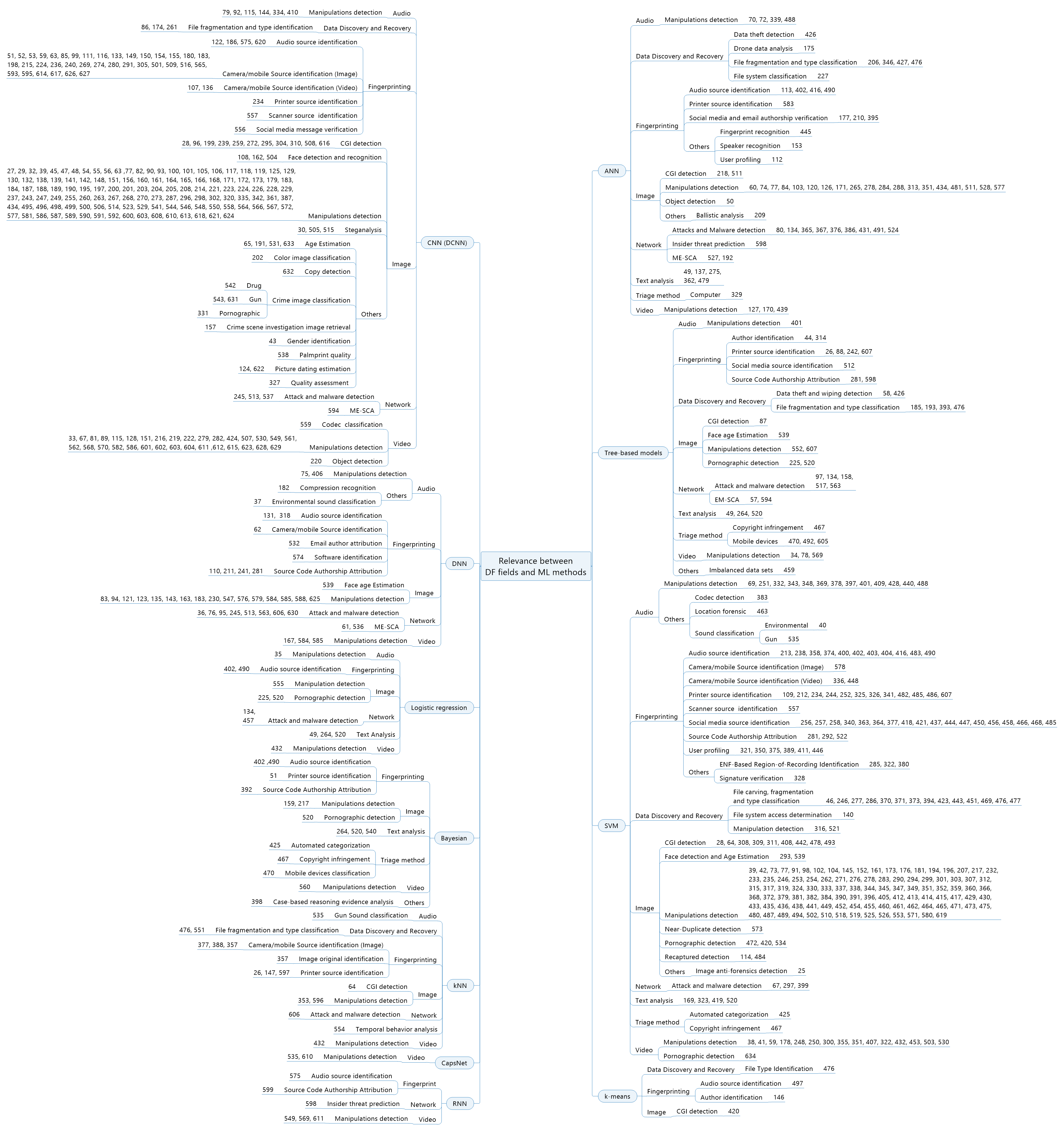}
    \caption{Mind map of the primary studies.}
    \label{fig:f10}
\end{figure}

Fig.~\ref{fig:f9} shows the main ML methods used in DF. Deep neural network (DNN) methods are placed in the category of DL with 14 repetitions in keywords. Furthermore, the Deep CNN (DCNN) keyword with eight repetitio-ns is placed in the CNN category. The 'others' group consists of the six methods that have ten or fewer repetitions in the authors' keywords: Bayesian, Logistic Regression (LR), KNN (k-nearest neighbour), LSTM (Long Short-Term Memory), CapsNet, and K-means. Based on the results obtained, deep learning, especially convolutional methods, plays a more meaningful role in digital and image forensics. It is considered a strength for digital forensics because of the more accurate results gained by CNN-based models. However, this may also be a weakness due to an increase in adversarial attacks. Using deep models in digital forensics has significantly grown since 2017. In 2021, about 53\% of the papers related to using CNN-based methods, and about 50\% were conducted in the image forensics domain.

In the following, some articles, mostly selected from Q1-ranked journals, are briefly reviewed to cover the main topics. A comprehensive grouping of all related works published from 2010 to 2021 is presented in Fig.~\ref{fig:f10}. As mentioned before, this mind map is based on the papers' context and covers all DF domains considered in papers. Considering the full text of the papers, this map shows that the CNN models have the most application in providing the DF approaches and frameworks, primarily in the image domain. Besides, it shows that the most significant concern in digital forensics is image manipulation. The dominant ML methods in this field have been CNN and Support Vector Machine (SVM). Another consideration is that SVM, tree-based, and Neural network-based (NN) models are effective in about all of DF categories, and K-means are used to a limited extent. One of the growing areas in data source identification is Social media source identification, and SVM techniques are used more. In the Triage models, traditional models such as Bayesian, Tree-based, and SVM are used, and DL models do not play a significant role in this regard.

\paragraph{Multimedia:}
With the development of mobile photography devices and editing software tools, the detection and identification of computer-generated images (CGI), recaptured images, and manipulated images have become the most severe challenges in multimedia forensics. Image manipulation includes a wide range of tampering and forgery, including splicing, copy-move, double compression, re-sampling, removal, resizing, sharpening, or smoothing operati-ons such as median filtering and Gaussian filtering. In the following, the main works in this field will be reviewed.

\begin{itemize}
\item \textbf{CGI and fake detection}

 \hfill \break
\textbf{image:}
Deep learning models, especially CNN-based models, have been successful in automatic multidimensional feature extraction~\cite{239, 272}, training, and classification for high accuracy recognition~\cite{96} due to their ability to obtain higher-order features. However, DL models have challenges. The performance of these models decreases in blind detection. For example, when training and testing data are generated with unknown computer graphics rendering tools, they deeply extract context from images without inferring any unique fingerprint. To solve this problem, Convolutional Traces analysis and feature extraction with Expectation-Maximization (EM) algorithm have had good results in the classification of fake images with SVM~\cite{64, 87}. Correspondingly, transfer learning~\cite{28, 304} can overcome two main DL models problems: 1) the need for a massive amount of training data and 2) overfitting on CNN. The reason is that the parameters of the pre-trained neural network (called source network), belonging to a particular task, are transferred to a new neural network (called target network), designed to solve somewhat similar tasks. A local-to-global strategy can also reduce the computational cost of images that are not the same size~\cite{508}. In this way, the CNN model will decide on local patches and the whole image in actual size through simple majority voting. Indeed, it crops several image patches with a fixed size in training, thus increasing the augmentation power of the training dataset.

SVM classification has been suggested to be applied to CGI detection in two aspects. The first is extracting histogram and multifractal spectrum features from residual images and regression model fitness features~\cite{493}. The second one is binary similarity measures of Photo Response Non-Uniformity (PRNU is a unique attribute in natural imag-es)~\cite{309, 408}. However, in combination with the DL methods, feature extraction can be done automatically. In NNs, using laplacian of gaussian, auto-correlation, and extreme learning machines have been suggested~\cite{511}. In fake colourized image detection~\cite{218}, extracting features from HSV colour space instead of RGB and NN training instead of SVM (due to low speed in large datasets and difficulty in choosing the correct kernel) can be a good solution. Detection in recaptured images can be done based on the difference in the number of pixels at the edges of the actual image and the recaptured image. This technique enables the detection of the image taken from the screens but presented as the original image. The same is possible for the hidden tampered images~\cite{114}.

 \hfill \break
\textbf{Video:} To distinguish fake videos from the original, extracting recompr-ession error can be a good criterion for detection using CNN-based models~\cite{424}. A fake detection scheme in~\cite{115} has been proposed for both video and audio with a convolutional recurrent neural network framework. In fake bitrate detection, SVM showed an excellent classific-ation accuracy~\cite{248}. In biometric authentication, video manipulation to fail authentication is more effortless compared to other authentication methods. An increase in AI-generated videos-based attacks leads to identifying the original videos from the tampered ones. In~\cite{81}, a hybrid CNN-LSTM model was proposed to improve the facial motion differences between fake and original videos.

 \hfill \break
\textbf{ Audio:}
Similar to other Multimedia forensics domains, data manipula-tion is one of the problems of audio forensics. The CNN-based method proposed in~\cite{334} determines whether an audio recording is recaptured or is genuine. The results showed that it has more robustness against ambient noise. Moreover, it can be detected appropriately for short 2-second clips.

\item \textbf{Manipulation detection}

 \hfill \break
\textbf{image:}
In the handcrafted-features-based methods, the learning and classifying steps cannot be simultaneously optimized~\cite{203}, as they are separate steps. In the CNN models, feature extractors and classifiers work automatically, and each feature map is generated at each network step. By performing a convolutional operation on the whole image and learning weights, features are extracted while training on a set of images. One of the essential advantages of these models is weight sharing. This capability makes them work faster and reduces data latency due to the computational resources of edge nodes while providing acceptable detection accuracy. Hence, weight sharing has been used in many works ranging from detecting forgery and locating it~\cite{29, 138, 267, 268}, based on the part of the image or the whole image and with any size~\cite{32}, to global manipulation detection and processing history detection~\cite{47, 118}. The results of~\cite{73} show analysis of colour filter array (CFA) using a trained SVM with original images and forged images, which leads to less computation time. Using the CNN models is recommended in small images.

Determining proper filter parameters can help in retrieving the manipu-lation history of an image. In~\cite{361}, authors suggested adding a transform layer to CNN and training it with frequency-domain features to identify template parameters of various spatial smooth filtering operations. Acc-urate locating of tampered and refined contours of tampered regions has been proposed in~\cite{529}. In some works~\cite{39, 82, 530}, the automatic feature extraction capability of the CNN models has been used. Then, in the classification step, other methods such as SVM are indicated to provide good results in two-class problems~\cite{39, 530} and the Forensic Similarity Graph approach proposed in~\cite{82}. However, the scheme presented in~\cite{54} extracts features based on median filter residual image creation using least-squares and employs CNN as the classifier. One of the CNN-based model features is the possibility of changing the kernel size. This feature improves the performance and is used to extract and integrate multi-scale features for copy-move tampering~\cite{296} and increase the reassembling rate estimation~\cite{101, 117}. Most deep-based methods pay attention to high-level features while there is much digital forensics evidence in low-level features. When these features are extracted, manipulations can be detected by training the network via a small dataset~\cite{83}.

Face images contain much personal information such as age, race, or even feelings. One of the challenges of the CNN models in image forensics is how to design a network for learning features through weak traces related to a specific manipulation~\cite{47, 132}. Because metadata is easier to manipulate in fake facial images, the RGB-based detection will work better~\cite{528}. In tamper detection, a determination is carried out if an image is derived from another one~\cite{183, 143}. Moreover, training a DNN with any pair of images in the face swap detection is also proposed~\cite{123}.

Compressed images are difficult to detect due to the influence of quality factors on multiple compressions, and the available methods work well under certain conditions~\cite{200}. In recent years, DNNs are suggested for JPEG compression manipulation detection~\cite{100, 167, 342}, non-aligned double JPEG compression detection~\cite{119, 496}, and resizing manipulation detection~\cite{184}. In~\cite{125}, features such as spatial, frequency, and compression increase detection performance. In the low-resolution image, which is a lossy compressed image that leads to the lack of statistical pixels to extract reliable features, filter layers and residual learning for Median filtering detection can be helpful~\cite{164}. SVM-based techniques are suitable to solve binary problems. Still, they are not appropriate for three-class issues, and most detection solutions are related to double compression. In a different work presented in~\cite{173}, the Triple JPEG Compressed Color Images detection was addressed. According to~\cite{181}, SVM obtains acceptable results for median filtering detection using an autoregressive model to extract features from bit-planes in uncompressed images. It also obtains promising results in uncompressed and JPEG compressed images based on the streaking effect.

In the field of splicing detection, Jinwei Wang et al.~\cite{93} proposed a CNN-based solution that uses a combination of YCbCr, edge, and PRNU features based on the weighting strategy and eliminates redundant information to obtain better results. In median filtering, a CNN-based model with an adaptive filtering layer was proposed in~\cite{514}, which is built upon the discrete cosine transform (DCT) domain. In social network image splicing detection, SVM is used for classification based on Texture Features~\cite{510}; SVM also utilizes Markov Features in QDCT and QWT Domains~\cite{253}. In~\cite{98}, using Vicinity Noise Descriptor with SVM has been proposed to solve the noise fluctuation problem in splicing zone detection.

For various types of motion or out-of-focus, blur image detection has been proposed using CNN-based model detection~\cite{188}. Furthermore, authors in~\cite{176} used multi-derivative grey level co-occurrence matrix (MGLCM) features to train SVM. Among image manipulations, sharp-ening is one of the most common techniques in most image editing tools~\cite{125}; however, its detection is a challenge in small images. To deal with this problem, using DCT-CHDMY features with SVM has been suggested~\cite{42}.

In Modified Contrast Enhancement-based forgery detection~\cite{284} and overlapping concurrent directional patterns~\cite{313}, NNs are suggested instead of SVM to obtain more robustness and accuracy. It should be mentioned that SVM had better results compared to NN in forgery detection of illumination component change using homomorphic image processing~\cite{196} and linear transformations such as rotation and resizing \cite{278}. Although the NN and DL models and conventional classifiers such as SVM have been used more in works, some of which are mentioned earlier, other techniques also presented good results. High accuracy was obtained in classification via KNN in splicing detection~\cite{353}, Naive Bayes (NB) in tampered JPEG image detection~\cite{159}, ensemble learning in seam carving forgery, and JPEG detection down-recompression~\cite{331}.

 \hfill \break
 \textbf{Video:}
Ease of manipulation risks the authenticity and integrity of digital videos, mainly when it is considered digital evidence in court. Unlike image data, video data contains temporal information. One of the problems with video manipulation detection is that there are no datasets with various tags for tampered videos to be large enough to be used in the ML techniques training. The CNN-based techniques can extract features and estimate various compression parameters such as quantification parameters, intra- or inter-frame type. Johnston and Elyan~\cite{128} deblocked the filter setting from good videos and trained a CNN model for manipulation detection. In relocated I-frames detection in double compressed videos, using a smaller CNN model for feature extraction and detection is appropriate to be embedded into mobile forensics devices~\cite{507}.

In compressed double videos, detecting abnormal frames in HEVC (High-Efficiency Video Coding) videos using NN has been proposed~\cite{439}. SVM classification is also considered in various fields due to its good compromise between computational complexity and detection accuracy \cite{503}. A comparison has been conducted in~\cite{432} among SVM, KNN, and logistic regression to detect deleted frames based on feature extraction from the bitstream and the reconstructed images. The results showed that SVM in CBR (constant bitrate) coding and LR in VBR (variable bitrate) coding has higher true positive (TP) rates compared to other techniques. Simultaneously, using KNN and linear discriminant analysis (due to their simplicity) and a three full connected multilayer perceptron (due to its accuracy) have better and more reliable performance in the classification of single and double compressed~\cite{170}.

 \hfill \break
 \textbf{Audio:}
 Voice biometrics is an approach that is increasingly used in speaker verification systems. Audio spoofing is one of the significant challenges in this field.~\cite{410}. Pitch shifting technique in audio editing and increasing or decreasing it is a prevalent voice manipulation to hide the true identity. Detecting a weakly pitch-shifted voice is one of the challenges in this field. The proposed CNN-based model in~\cite{92} has been able to overcome it. D. Luo et al.,~\cite{339} used NN in double compressed AMR audio detection via a stacked autoencoder to learn the optimal features of audio waveforms. For higher efficiency, they extracted compressed-domain speech features instead of decoded speech waveforms from encoded AMR files. However, the accuracy in~\cite{69} has been higher because of using SVM classification. Reis et al.~\cite{343} utilized electrical network frequency to detect and identify tampered audio recordings by detecting abnormal variations.

\hfill \break
\item \textbf{Text analysis}
Due to daily user activities such as email, web browsing, word processing, etc., textual evidence is vital. Much of this evidence can be in the unallocated space, or the file's signature has been overwrit-ten. Lang Beebe et al.~\cite{419} employed SVM to provide ranking algorith-ms for the search string. In Uyghur Web Text Classification~\cite{323}, SVM also presented good results. In automating the analysis of text and log chats analysis, the ML techniques help investigators analyze and identify conversations containing sexually inappropriate subjects~\cite{49}. In the child sexual abuse and the identification of Unknown Criminal Media in P2P networks, SVM performed better than Naive Bayes and LR in text and image classification~\cite{520} regarding the F-score. Identifying the source of evidence is essential in the investigation of online communications on social networks and email. In~\cite{540}, a ML-based framework and natural language processing has been developed. The results indicated that random forest (RF) has the best average precision in classification compared to the Decision tree (DT), NB, AdaBoost, LR, and SVM.

\hfill \break
\item \textbf{Other multimedia fields}
A SVM-based classification tool has been recommended in~\cite{534} to reduce costs and speed up the investigation. They used a parallel software architecture for photo and video multimed-ia evidence classification easily and instantly. If the results are positive and there is evidence, the device should be taken to the laboratory for further analysis. In face detection and Sketch Synthesis (FSS), drawing the suspect face showed promising results by applying the IEHO algorithm to hyperparameter optimization in the proposed CNN-based model~\cite{108}. Although existing DL modelling tools work well on adult age estimation, they do not perform well on underage subjects. Authors in~\cite{65, 531} improved a DL-based approach, especially in child sexual abuse investigations. Their scheme performed well by obtaining a large dataset with balanced and valid class labelling. In gender identification, good results were reported in CNN-based gender identific-ation of handwriting by extracting hierarchical page features and word features~\cite{43}.

In video object detection, there is footage recording quality dependency. The low quality of the recording will reduce the reliability of the evidence collected for presentation to the court. In object detection in the footage, the colour of a particular object is not consistent for reasons such as changing the quality of video or changing light, etc., all of which make identification difficult. In this case, the DL methods will be more efficient in identifying the image object~\cite{220}. Due to the high noise in the circumference, environmental sound classification (ESC) is one of the most challenging problems for DF and ML. Several solutions based on DNN~\cite{37} and SVM~\cite{40} were proposed to solve these problems. In sound-based gun model identification, which is complex and costly to do manually, the KNN classifier showed better accuracy than SVM~\cite{535}.
\end{itemize}

\paragraph{Fingerprinting}
In identifying a suspected source, which could be a printer, a digital camera, malicious code writers, or even the way people behave, often there are clues in relevant evidence that help investigators. Footprints can be left in various data such as images, audio, video documents, etc. The ML methods can help investigators to extract and analyze these footprints more accurately and quickly.

\begin{itemize}
\item \textbf{Camera or phone source identification}

\hfill \break
\textbf{image:}
The issue of device source identification based on their fingerpr-ints is a growing topic in digital forensics~\cite{33}. Source camera identificat-ion is a serious issue in digital forensics areas such as copyright and property~\cite{305}, including camera manufacturer identification and camera model identification (CMI). The DL models like CNN can automatically extract features and identify multiple camera models~\cite{51, 149, 198, 509, 516}. They have also been used in pre-processing tasks to remove scene content~\cite{116}, which severely hides the camera fingerprints. Distinguish-ing between images taken directly by a camera or downloaded from social networks is challenging in image forensics. In this field, using SVM classifiers is taken into consideration~\cite{256, 257, 258, 340}. In source camera identification without knowing the models, using the KNN classifier and a self-training strategy showed better results than a binary SVM and a multi-class SVM~\cite{377}. KNN is also used in multimedia phylogeny~\cite{357} to find abusive or original content publishers.

In uploaded images to messenger applications or social networks, remai-ning footprints can be extracted to help identify the image source. In this regard, a CCN-based camera detection on a shared document has been proposed~\cite{501}. One of the challenges in messengers is that filters applied to images can cause sensor pattern noise (SPN). This can cause camera detection methods not to work correctly. A CNN-based solution in~\cite{180} has been suggested for messenger apps like Whatsapp to identify these filters. Another challenge is low accuracy with a limited set of labels for training. In~\cite{62}, the deep siamese network significantly increased the classification accuracy. The RF has been proposed to identify image sources from social networks like Facebook, Twitter, and Flickr~\cite{512}.

\hfill \break
\textbf{Video:}
The supervised ML algorithms effectively detected the video editing tool used in social networks and messengers, and higher perform-ance was obtained by random forest~\cite{34}. In~\cite{84}, a container-based method has been proposed to identify software and operating systems that manipulate videos on social networks. They also employed a decision tree to explain decisions.

\hfill \break
\textbf{Audio:}
Identifying a mobile device based on taken images is an approach that has been considerably used in the literature~\cite{51, 62, 116, 149, 180}, as discussed earlier. Another method that can be employed to identify cell phones is microphone identification. Using directly extracted features from audio signals is a common way. It works well in identifying different brands, but it fails to distinguish different brand models. In~\cite{122}, a CNN-based model identified cell phones through their built-in microphone by extracting features from different parts of the spectrogram of multiple streams~\cite{122}. In addition to identifying different brands, it also identifies different models of the same brand. Using NN with device noise feature~\cite{113} and sparse representation-based classification method~\cite{497} have their benefits in source recording device recognition.

One of the challenges in audio forensics is recognising VoIP calls as caller-ID that can be easily manipulated in such calls. Among ML-based proposed methods for this problem, SVM and NN showed good accuracy and robustness~\cite{490}; however, they can be used to identify a specific source device. Hence, a CNN model was suggested in~\cite{186} based on temporal and spectral domain features.

\hfill \break
\item \textbf{Printer source identification}
Printer or scanner source identification is vital in matters such as copyright ownership or document authent-ication manipulation. In recent years, using SVM has been discussed in various areas such as printer source identification of a text document~\cite{109, 212, 325}, printer source identification for both text and image using image processing and microscopic image techniques~\cite{252}, and data exploration methods~\cite{244}. Furthermore, the method composed of Naive Bayes classifiers, KNN, and RF has been proposed in~\cite{26} as a feature extraction method (to extract Speeded Up Robust and Oriented Fast Rotated features). Random forest algorithm presented better performance than SVM due to feature importance that allows explaining a method with less run-time and no need for kernel and parameterization adjustments~\cite{242}.

\hfill \break
\item \textbf{Authorship attribution and Profiling}
In digital forensics, the details of the document author, such as identity, demographic information of the authors, etc., are significant~\cite{44}. Authorship verification is trying to identify whether the authors of the two documents are the same or not. This is a challenging issue due to the shortness of the messages on social networks~\cite{210}. In programming, Source Code Authorship Attribution can be necessary for digital forensics in some ways, including developer privacy and the importance of identifying malicious code programmers. To identify three programming languages like C++, Java, and C\#, the accuracy of the DL-based technique among 100 programmers was reported to be 97.34\%~\cite{110}.

The results of~\cite{211} showed that using syntax tree and deep learning versus Program Dependence Graph with Deep Learning has a lower complexity cost. Because the time and rhythm of people typing can have a specific pattern, keystroke-based identification techniques are used as the biometric tool to identify individuals. The lack of a dataset with real keystroke dynamics is one of the problems in this field. Authors in~\cite{112} created a new keystroke dynamic dataset. They also showed that classification with radial basis neural networks was more accurate than the other ML models.
\end{itemize}

\paragraph{Network}

\begin{itemize}
\item \textbf{Attack and malware detection}
In detecting the source of cyber-attacks, the high accuracy and low false alarms rate are important parameters. Among the DNN models, the multilayer perceptron algorit-hm~\cite{36} showed promising results in identifying attacks for high-volume data. Among the NN-based models, Neuro-Fuzzy-based techniques~\cite{365, 376, 491, 537} can be used in network traffic analysis to extract accurate and interpretable data. The Fuzzy Min-Max NN model~\cite{524} with online adaptation and online learning capability can add new classes or modify existing classes without the need for retraining.

Autoencoders do not learn patterns that they have not seen before, so multi-autoencoders have been suggested in~\cite{95}. A One-Class SVM model and a semi-supervised model can be trained using the appropriate core function, only with data from a normal class (without attack). They can then be utilized to classify normal events from abnormal~\cite{66, 297}. In an under-attack situation such as critical infrastructures, various critical processes are executed, each of which can have its attack. By applying the appropriate SVM settings for each process, the specific attacks can be more accurately identified.

Deep learning models are complex in detecting and analyzing malware and require much training and pre-processing time, which is unsuitable for critical infrastructure environments. The NN models can be faster options~\cite{513, 386}. Specifically, CNNs are Shift invariant, meaning the CNN models do not require the expertise of an investigator. In the models that rely on signatures and malware behaviour, raw static byte code as input~\cite{245} can solely solve time-consuming and high computational resource consumption challenges. For detecting malicious web pages, authors in~\cite{76} proposed a DNN-based tool, which reported 99.8\% accuracy. Among tree-based models, Boosted tree methods can provide investigators with a simple approach that detects malware faster than an antivirus. Likewise, among various ML techniques, the DT can perform better in malware family classification due to having a comprehensive analysis nature and effective in reducing false alarm rates~\cite{517}. On the networks with Windows operating systems, extracting features from the registry to train Boosted tree~\cite{134} reported good results compared to NN, LR, and DT. The use of LR in identifying the reason for scavenging adversaries accessing data has been suggested in IoT environments~\cite{457}. In the fields such as cyber threat intelligence (CTI)~\cite{158}, RF can have good results in malware classification thanks to reducing the total token count.

\hfill \break
\item \textbf{Attack and malware detection}
With the expansion of the IoT, the devices used in these networks could be a potential source of evidence for Digital forensics. However, the diversity of devices and the lack of a standard interface are some of the problems facing DF investigators. Because of the limited interface of smartphones, retrieving valuable forensics information is a challenging task. Recently, researchers have employed EM-SCA attacks to gather information from IoT devices in a forensically sound way. However, the successful implementation of this attack requires knowledge and equipment that most investigators do not have. To solve this challenge, evidence collection and analysis can be done automatically~\cite{536}.

Although the analysis of the massive amount of collected EM traces is complicated, the efficiency of the analysis can be increased if a subset of frequency channels with sufficient information can be obtained. Using the RF classifier~\cite{57} due to its fastness and high accuracy can be suitable for identifying information-leaking frequency channels from EM side-channel data with high dimensions and selecting features (time domain and frequency domain properties). In the DL models, the large volume of collected data can increase the accuracy of data analysis by receiving an input vector with a larger size. However, this data is not directly suitable for the input and training of the DL models and requires pre-processing. In a recurrent neural network (RNN), the LSTM architecture is appropriate to identify the patterns that occur in time series data. With the Fast Fourier Transform (FFT) vector, LSTM can be trained to distinguish elliptic curve cryptography (ECC) operations from other software activities~\cite{527}.
\end{itemize}

\paragraph{Data Discovery and Recovery}
In data carving and fragmentation, identifying the data type is essential because it helps identify the type of crime. Large-scale file fragment type identification is one of the challenges in this field. Implicit extraction of features with a CNN model has better performance and run-time than similar works~\cite{86}. Another challenge is related to the fragmented JPEG files whose metadata is lost intertwined with non-JPEG files in the scanned area. Using extreme learning~\cite{206}, better accuracy and timing have been reported to classify and identify JPEG than non-JPEG. Using SVM in this area has also been considered, which consists of improving the accuracy of classification of file systems accessed during a digital crime~\cite{140}, improving data type identification~\cite{443}, and file fragment classification~\cite{277, 469}. Based on n-gram analysis~\cite{533}, SVM-based approaches performed better in file type identification than NN. However, their scalability is still a challenge.

In data theft detection, regression tree~\cite{426} performs better than neural networks. Moreover, in data wiping, where files are securely deleted, one of the abuses is related to eliminating evidence. Random forest with high accuracy~\cite{28} can help investigators identify what data was deleted and what tools were used. Because some devices have built-in encryption algorithms for security reasons, the EM-SCA process becomes more complex. There is also a need to identify and classify encryption algorithms. Neural networks with FFT vector~\cite{192} and DNN with amplitudes as input~\cite{61} are used effectively in this field.

\section{Future research directions of ML application in DF}
This study aims to provide an exhaustive review of the ML applications in DF over the past decade. We present the details of ML applications in various domains of digital forensics. Based on the results of this survey, we present the following research directions of machine learning for digital forensics that are worth further investigation:

\begin{itemize}
\item \textbf{} The lack of a standard taxonomy and ontology for a The complete classification of digital forensics domains based on the scope of their applications and the type of evidence is one of the crucial issues the authors faced in this study. For example, in the discussion of image source identification, different researchers have used different terms. Some have included it in printer forensics, camera forensics (device forensics), and some in image forensics. Different categories such as e-mail forensics, network forensics, or text forensics were used in cases related to identifying the author of a text.

\item \textbf{} This work is a systematic review to identify the state of ML methods in digital forensics. The results showed that the CNN-based models are the dominant method in studies. However, researching the technical aspect, identifying the level of security of using these methods, and examining the effect of adversarial attacks on the proposed methods can significantly help increase the security of the CNN-based methods.

\item \textbf{} Adjusting many parameters in the DL models and defining the usefuln-ess of layers in some applications are among the problems raised in the related works. One of the essential needs, especially in image forensics, can be related to collecting the relevant settings in the investigations and comparing the results to identify the best settings in each digital forensics application.

\item \textbf{} Except for image and video forensics, using the DL models in other domains is relatively less. Consequently, further attention and research are required to take advantage of the DL capacity in other digital forensics domains.

\item \textbf{} The focus in using ML in the DF process is on evidence acquisition and detection. Using ML methods in the evidence reconstruction and analysis phase could be valuable in the future of DF research.
\end{itemize}

\section{Conclusion}
In recent years, the growth of digital data and the increase of tools to facilitate digital crime showed that traditional digital forensics methods are no longer responsive. There is a need to automate processes to increase speed and accuracy in investigating and analyzing the results. In particular, ML techniques have been used for a long time in various fields, and their features have also been growing in digital forensics. Having a general view of the conducted investigations is necessary to identify the leading ML methods in DF and the affected DF domains, discover existing gaps, and clarify the future scientific path for researchers. Thus far, studies and surveys have been conducted, but these studies have some limitations. They have focused on a specific DF domain or only have examined the specific ML methods. In this paper, for the first time -as far as we know-, we review the last ten years of research related to using ML in Digital forensics. Furthermore, colour keywords-based visual analysis and a comprehensive Mind Map were provided to identify the applications of ML methods in DF domains. It can be used as a general map for digital forensics researchers.

The growing trend of using ML in DF indicates that it effectively improves the DF process, and there are still open research areas. As the meta-analysis of this study showed, CNN models have found an important place in DF. Due to the increase in attacks on these models, it is appropriate to consider the security of these models in DF in future research. The number of studies related to DL models other than image and video was significantly lower. Due to the high performance of these models, more research on their application in other DF domains will be valuable. Another important and exciting topic in future research is expanding ML methods in the investigation process and more DF phases and paying attention to explainable ML in digital forensics.

 %%\bibliographystyle{elsarticle-num}
 %%\bibliography{cas-refs}





% \end{thebibliography}
\end{document}